\documentclass[3p,times,twocolumn]{elsarticle}

\usepackage{ecrc}

\volume{X.Y}

\firstpage{1}

\journalname{Nuclear Instruments and Methods in Physics Research A}

\runauth{}

\jnltitlelogo{Nuclear Instruments Methods}

\usepackage{amssymb}
\usepackage{amsmath}
\usepackage{upgreek}

\usepackage{lineno}

\usepackage[figuresright]{rotating}

\usepackage{scalefnt}

\newcommand{\mm}{\rm{\upmu m}}
\newcommand{\ms}{\rm{\upmu s}}

\begin{document}

\begin{frontmatter}



\dochead{}

\title{Performance of a Full-Size Small-Strip Thin Gap Chamber Prototype for the ATLAS New Small Wheel Muon Upgrade}


\author[7]{A.~Abusleme}
\author[2]{C.~B\'elanger-Champagne}
\author[1]{A.~Bellerive}
\author[8]{Y.~Benhammou}
\author[1]{J.~Botte}
\author[8]{H.~Cohen}
\author[8]{M.~Davies}
\author[12]{Y.~Du}
\author[3]{L.~Gauthier}
\author[1]{T.~Koffas}
\author[6]{S.~Kuleshov} 
\author[2]{B.~Lefebvre}
\author[12]{C.~Li}
\author[9]{N.~Lupu}
\author[10]{G. Mikenberg}
\author[4]{D.~Mori}
\author[7]{J.~P.~Ochoa-Ricoux}
\author[5]{E.~Perez Codina}
\author[5,11]{S.~Rettie}
\author[2]{A.~Robichaud-V\'eronneau}
\author[6]{R.~Rojas}
\author[10]{M.~Shoa}
\author[10]{V.~Smakhtin}
\author[4]{B.~Stelzer}
\author[5]{O.~Stelzer-Chilton}
\author[6]{A.~Toro} 
\author[4]{H.~Torres}
\author[6]{P.~Ulloa}
\author[2]{B.~Vachon}
\author[6]{G.~Vasquez}
\author[9]{A.~Vdovin}
\author[5]{S.~Viel}
\author[7]{P.~Walker}
\author[1]{S.~Weber}
\author[12]{C.~Zhu}

\address[1]{Carleton University, 1125 Colonel By Drive, Ottawa, ON, K1S 5B6, Canada}
\address[2]{McGill University, Ernest Rutherford Physics Bldg., 3600 Rue University, Montreal, QC, H3A 2T8, Canada}
\address[7]{Pontificia Universidad Catolica de Chile, Vicu\~na Mackenna 4860, Santiago, 7820436, Chile}
\address[4]{Simon Fraser University, 8888 University Drive, Burnaby, BC, V5A 1S6, Canada}
\address[9]{Technion - Israel Institute of Technology (IIT), Haifa, 32000, Israel}
\address[8]{Tel-Aviv University, Ramat Aviv, Tel Aviv, 69978, Israel}
\address[5]{TRIUMF, 4004 Wesbrook Mall, Vancouver, BC, V6T 2A3, Canada}
\address[6]{Universidad Tecnica Federico Santa Maria, Casilla 110-V, Avda. Espana 1680, Valparaiso, Chile}
\address[3]{Universit\'e de Montr\'eal, C.P. 6128, Succ. centre-ville, Montr\'eal, QC, H3C 3J7, Canada}
\address[11]{University of British Columbia, 6224 Agricultural Road, Vancouver, BC, V6T 1Z1, Canada}
\address[12]{Shandong University, Jinan, Shandong, China}
\address[10]{Weizmann Institute of Science, Rehovot, 76100, Israel}

\begin{abstract}
The instantaneous luminosity of the Large Hadron Collider at CERN will be increased up to a factor of five with respect to the present design value by undergoing an extensive upgrade program over the coming decade. The most important upgrade project for the ATLAS Muon System is the replacement of the present first station in the forward regions with the so-called New Small Wheels (NSWs). The NSWs will be installed during the LHC long shutdown in 2018/19. Small-Strip Thin Gap Chamber (sTGC) detectors are designed to provide fast trigger and high precision muon tracking under the high luminosity LHC conditions. To validate the design, a full-size prototype sTGC detector of approximately 1.2 $\times$ $1.0\, \mathrm{m}^2$  consisting of four gaps has been constructed. Each gap provides pad, strip and wire readouts. The sTGC intrinsic spatial resolution has been measured in a $32\, \mathrm{GeV}$ pion beam test at Fermilab. At perpendicular incidence angle, single gap position resolutions of about $50\,\mm$ have been obtained, uniform along the sTGC strip and perpendicular wire directions, well within design requirements. Pad readout measurements have been performed in a $130\, \mathrm{GeV}$ muon beam test at CERN. The transition region between readout pads has been found to be 4\,mm, and the pads have been found to be fully efficient.
\end{abstract}


\end{frontmatter}



\section{Introduction}
\label{sec:intro}
The motivation for the luminosity upgrade of the Large Hadron Collider (LHC) is to precisely study the Higgs sector and to extend the sensitivity to new physics to the multi-TeV range. In order to achieve these goals the ATLAS experiment \cite{ATLAS} has to maintain its capability to trigger on moderate momentum leptons under more challenging background conditions than those present at the LHC during Run-1 and Run-2.
For the Muon Spectrometer (MS) \cite{ATLAS_MS}, 
such requirements necessitate the replacement of the forward muon-tracking region called the muon Small Wheel, with new detectors capable of triggering and precision tracking simultaneously. The New Small Wheel (NSW) upgrade \cite{NSW_TDR} is designed to cope with the high background rate that is expected at luminosities between $2 - 7 \times 10^{34}\, \mathrm{cm}^{-2}\mathrm{s}^{-1}$ during Run-3 and the high luminosity LHC (HL-LHC) runs \cite{HL-LHC}. 

\begin{figure}[h!]
\centering
\includegraphics[width=0.9\linewidth]{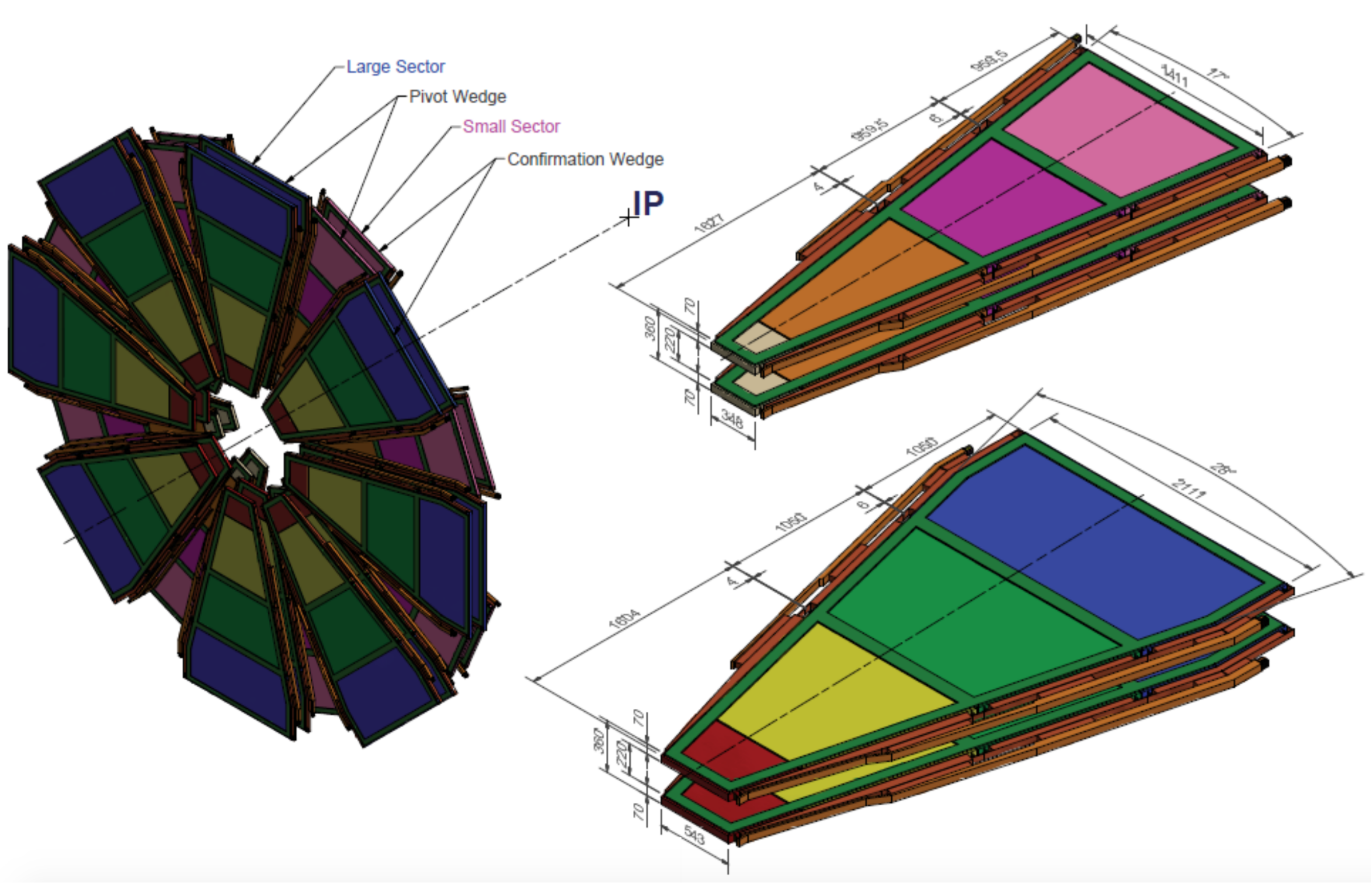}
\caption{Schematic diagram of the small and large sectors that make up the Muon New Small Wheel. Each sector consists of two quadruplets of sTGC with eight micromegas detection planes in between.}
\label{fig:sTGC_prod}
\end{figure}

Small-Strip Thin Gap Chambers (sTGC) have been selected as one of the two detector technologies that will be used for the NSW upgrade along with micromegas detectors. Fig. \ref{fig:sTGC_prod} shows a schematic diagram of the NSW. The NSW includes eight detection planes (layers) of sTGC arranged in two quadruplets and eight planes of  micromegas. The precision reconstruction of tracks for offline analysis requires a spatial resolution of about $100\,\mm$ per sTGC layer, and the track segments have to be reconstructed online for triggering purposes with an angular resolution better than $1\,\mathrm{mrad}$. A large collaboration has been established to construct these devices and is composed of members from Canada, Chile, China, Israel and Russia. 
These precision requirements are challenging to achieve, therefore beam test experiments have been performed  to qualify the sTGC assembly procedure.

\section{sTGC Technology}
\label{sec:technology}
\begin{figure}[t!]
\centering
\includegraphics[width=0.7\linewidth]{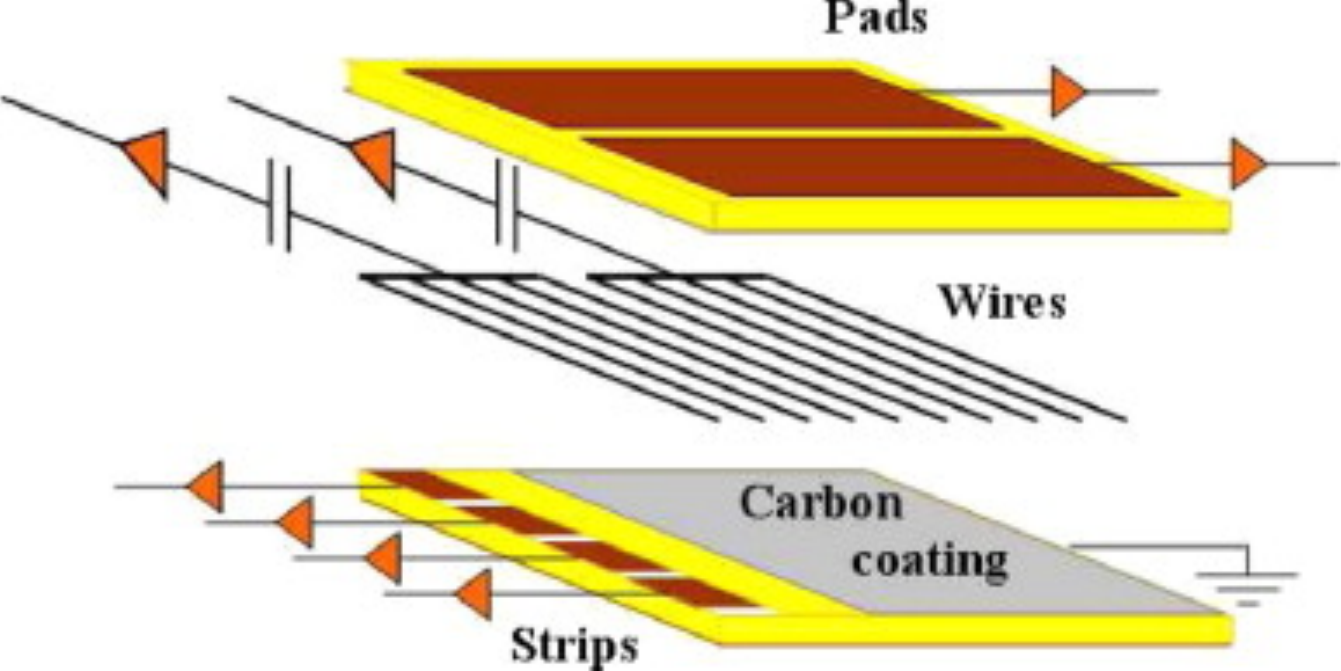}
\caption{Schematic diagram of the basic sTGC structure.}
\label{fig:sTGC_tech}
\end{figure}

The concept of Thin Gap Chambers (TGC) was developed in 1983 \cite{TGC} and then used at the OPAL experiment and for the ATLAS end-cap muon trigger system.
The basic sTGC structure is shown in Fig. \ref{fig:sTGC_tech}. It consists of an array
of $50\,\mm$ diameter gold plated tungsten wires held at a potential of $2.9\,\mathrm{kV}$, with a $1.8\,\mathrm{mm}$ pitch, sandwiched between two cathode
planes located at a distance of $1.4\,\mathrm{mm}$ from the wire plane. The cathode planes are made of a graphite-epoxy mixture with a typical surface resistivity of 100 or $200\,\mathrm{k}\Omega/\square$ sprayed on a 
100 or $200\,\mm$ thick G-10 plane for the inner and outer chambers, respectively.
Behind the cathode planes on one side of the anode plane there are copper strips for precise 
coordinate measurements that run perpendicular to the wires and on the other side of the anode plane 
there are copper pads used for fast trigger purposes.
The copper strips and pads act as readout electrodes. The pads cover large rectangular surfaces on a 1.5 mm thick printed circuit board (PCB) with the shielding ground on
the opposite side. The strips have a $3.2\,\mathrm{mm}$ pitch, much smaller than the strip
pitch of the ATLAS TGC\footnote{The strip-pitch of the TGC varied between $150\,\mathrm{mm}$ to $490\,\mathrm{mm}$}, hence the name 'small-strip TGC' for this technology.

Each sTGC quadruplet consists of four pad-wire-strip planes shown in Fig. \ref{fig:sTGC_tech}.
The pads are used through a 3-out-of-4 coincidence to identify muon tracks approximately pointing back to
the interaction point. They are also used to define a region of interest that determines which group of strips needs to be read out in order to obtain
a precise position measurement in the precision coordinate, for the online event selection. The azimuthal
coordinate of a muon trajectory is obtained from the wires readout. The operational gas is a mixture of 55\% CO$_2$ and 45\% n-pentane.
There are six different sizes of sTGC quadruplets, three for each of the large and small sectors.
As shown in Fig. \ref{fig:sTGC_prod}, all have trapezoidal shapes with surface areas between 1 and $2\,\mathrm{m}^2$.

\section{Construction of a Large sTGC Prototype}
\label{sec:prototype}
A challenge in the construction of large area multi-layer particle detectors
is to achieve high precision alignment of the readout strips across layers. 
The required accuracy in the position and parallelism of the precision strips between planes is $40\,\mm$. This precision is achieved by mechanical machining. The 
readout strips for an sTGC plane are machined together, in one step, with brass inserts which can be externally referenced. 
The cathode boards are glued together, separated by chamber walls at the periphery of the boards as well as $7\,$mm wide T-shaped wire supports and spacer buttons in approximately 20\,cm intervals. The resulting individual chambers are glued together, separated by an especially-machined frame with 
a honeycomb structure over the entire surface of each chamber, which is smaller by $100\,\mm$ than the gap between chambers.
The glue serves as a filler to compensate for small deviations in the thickness of the PCB material.
The gluing procedure makes use of the fact that the various sTGC  layers can be positioned with respect to each other with high accuracy, 
using the external brass inserts attached to an external precision jig on a marble table. 

In the spring of 2014, the Weizmann Institute of Science in Israel built the first full-size sTGC quadruplet detector of dimensions $1.2 \times 1.0\,\mathrm{m}^2$.
This prototype consists of four sTGC strip and pad layers and is constructed using the full 
specification of one of the quadruplets to be used in the NSW upgrade (the middle quadruplet of the small sector).

\section{Readout Electronics of the sTGC Prototype}
\label{sec:readout}
A versatile application-specific integrated circuit (ASIC) is being
developed to read out the pads, strips and wires of the sTGC detectors.  
The first prototype version of this ASIC
(the so-called VMM1~\cite{VMM1} chip), was used to read out pads and strips at the Fermilab 
and CERN beam tests described below. This is the first time the VMM1
ASIC was used to measure the performance of a full-size sTGC
prototype detector.

The VMM1 analog circuit features a charge amplifier stage followed
by a shaper circuit and outputs the analog peak value (P) of the signal.  The
readout of the ASIC is zero suppressed and thus 
only peak values of channels with signals above a predefined threshold
are read.  The VMM1 may be programmed to also provide the input
signal amplitude of channels adjacent to a channel above threshold
(neighbour-enable logic).  The VMM1 chip has the ability to read out
both positive (strips, pads) 
and negative (wires) polarity signals, on 64 individual
readout channels.  
The shaping amplifier features an adjustable peaking time (25, 50, 100, $200\,\mathrm{ns}$) and is realized using the delayed dissipative feedback architecture which offers lower noise and higher dynamic range. The VMM1 also features an output baseline stabilizer circuit and the gain is configurable (0.5, 1.0, 3.0 and $9.0\,$mV/fC).
An internal global DAC and a $1\,\mathrm{pF}$ calibration capacitor provide
the ability to send test signals of different selectable charges to each
individual readout channel.
Finally, the VMM1 has an analog monitor output which
can be programmed to output the analog waveform after the shaping
stage of any of the 64 channels.

The analog signal amplitude is designed to be proportional to the
input charge. An estimate of the input signal
charge is therefore obtained by subtracting each channel baseline from its readout
analog peak value.  The average VMM1 channel baseline is 
approximately 180\,mV with a channel-by-channel baseline variation of
up to $\pm$3\% around the average baseline value.

The analog peak values of channels above threshold (and possibly
adjacent channels) are digitized by a 13-bit ADC on a separate custom
data acquisition card providing input/output Ethernet interface. Both
the configuration of the VMM1 and digitized readout of the channels
peak values are transmitted over Ethernet through the custom data acquisition card.

The precise position of a charged particle traversing an sTGC gas
volume can be estimated from a Gaussian fit to the measured charge
on adjacent readout strips (referred to as strip-clusters from here on). Given the strip pitch of 3.2\,mm and sTGC geometry, charges are
typically induced on up to five adjacent strips. The spatial sampling of
the total ionization signal over a small number of readout channels means
that a precise knowledge of each individual readout channel baseline
is necessary in order to achieve the best possible measured spatial
resolution.  The baseline of each individual readout channel was
measured by making use of the neighbour-enabled logic of the VMM1 and its
internal calibration system.  Test pulses were sent on one readout
channel with the neighbour-enabled logic on, and baseline values were
obtained by reading out the analog peak values of the two channels
adjacent to the one receiving a test pulse.  Baseline values for each
individual readout channel were measured and observed to be stable
as a function of time (to better than 1\%).

\section{Position Resolution Measurements at Fermilab}
\label{sec:fermilab}
The main goal of the beam test experiment at Fermilab was to determine the position resolution of the first full-size sTGC prototype detector. An external silicon pixel tracking system was employed to precisely characterize the sTGC performance and aid in the determination of the intrinsic spatial resolution. Previous measurements of the resolution of other sTGC prototypes, including determining the dependence of the resolution on the track incidence angle are described in \cite{sTGC_res1,sTGC_res2,sTGC_res3}.
\subsection{Experimental Setup at Fermilab}
\label{sec:expsetup}
\begin{figure}[t!]
\centering
\includegraphics[width=0.7\linewidth]{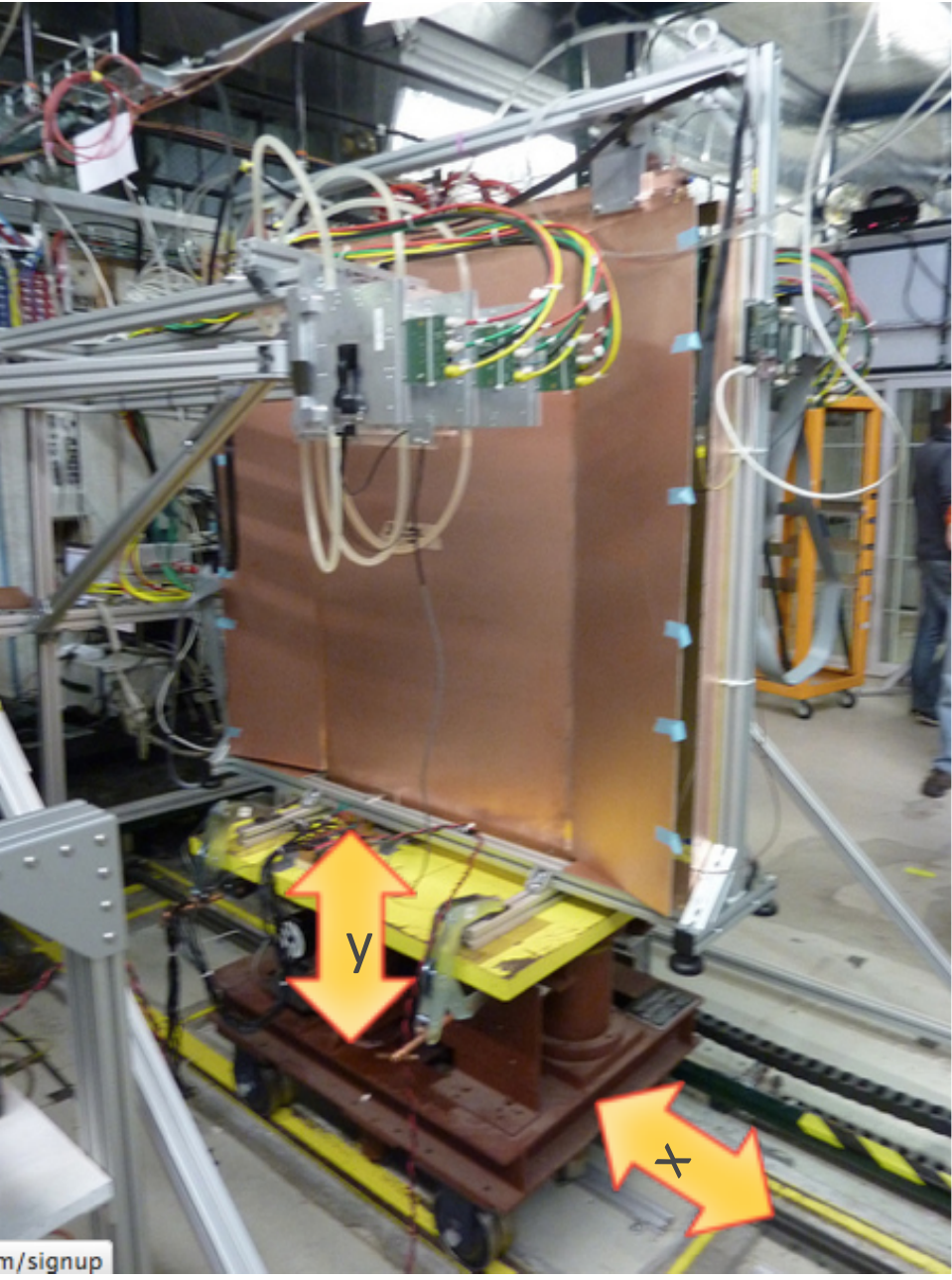}
\caption{Experimental setup of the beam test at the Fermilab Test Beam Facility. The sTGC prototype detector, inside a copper Faraday cage, was mounted on a 
motion table. Two arms of the EUDET silicon pixel telescope were positioned before and after the sTGC detector.}
\label{fig:exp_setup}
\end{figure}
In May 2014, the full-size sTGC prototype was tested with a 32\,GeV pion beam at
the Fermilab Test Beam Facility. 
The beam intensity was approximately 4,000 particles per spill and corresponded to a particle rate of about 1\,kHz on average with a 1\,cm$^2$ beam spot size.

The beam enclosures were outfitted with laser systems that allowed the determination 
of the beam location in the transverse plane of the experimental setup.
As shown in Fig.~\ref{fig:exp_setup}, the sTGC detector was mounted onto a motion table that was controlled remotely with a 1\,mm precision.
The particle beam was along the z-axis, while 
the sTGC detector was moved in the x and y-directions in order to test different areas of the detector.
\begin{figure}[b!]
\centering
\includegraphics[width=0.9\linewidth]{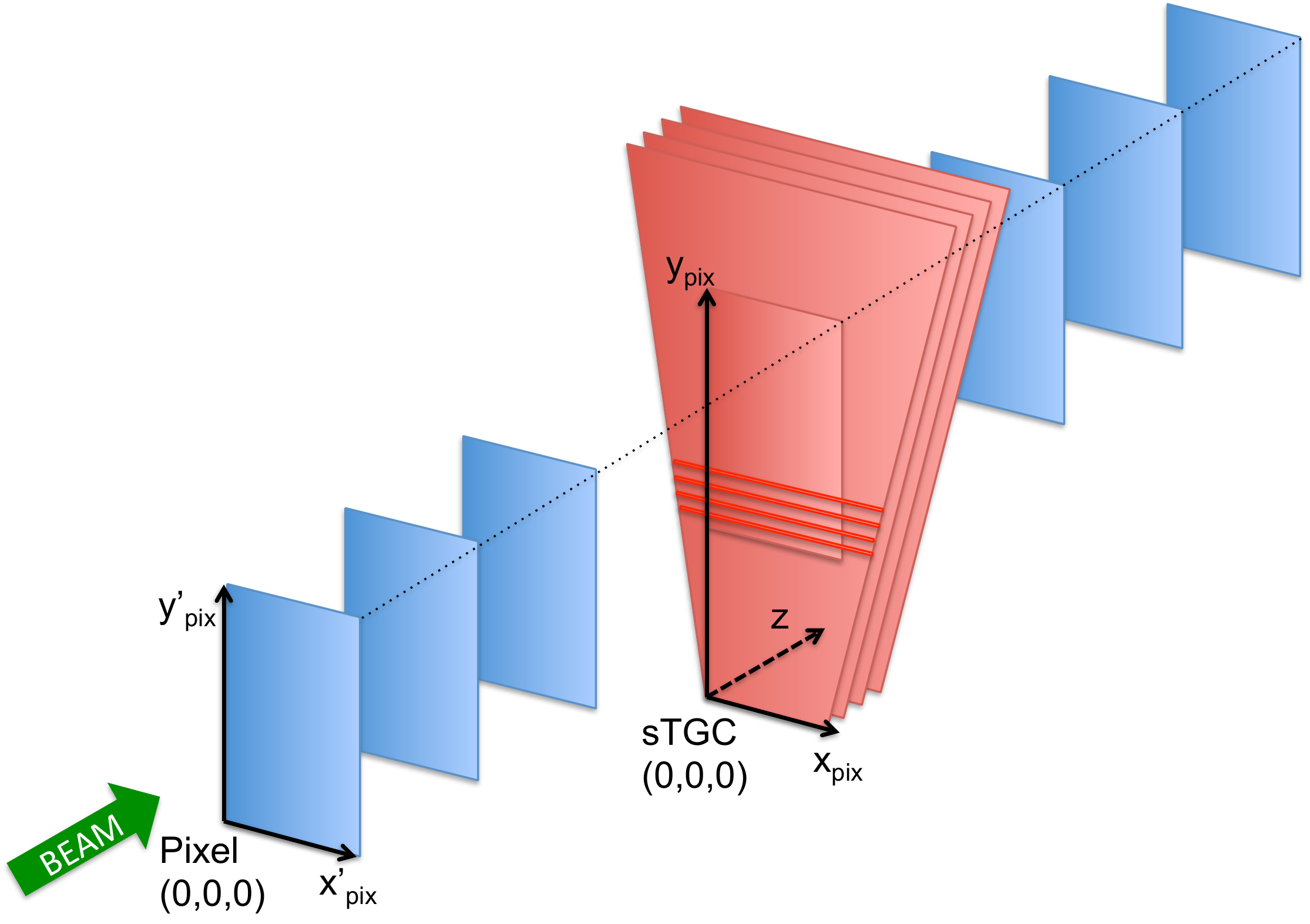}
\caption{Schematic diagram of the experimental setup at Fermilab and coordinate systems used. Three layers of silicon pixel sensors are positioned before and after the sTGC detector. The dimensions are not to scale.}
\label{fig:schematic_setup}
\end{figure}

The sTGC was positioned between the two arms of the EUDET pixel telescope~\cite{eudet} as shown schematically in Fig. \ref{fig:schematic_setup}. The telescope 
consisted of six Minimum Ionizing MOS Active Pixel Sensor (Mimosa) planes that have a 224\,mm$^2$ active area. Each arm consisted of three sensor
planes. For each arm, the distance between sensor planes was about 15\,cm. The two arms were separated by 64\,cm. 
The Mimosa of the EUDET telescope
offered excellent performance in terms of intrinsic resolution, material budget and readout 
electronics. The Mimosa26~\cite{mimosa} $18.4\,\mm$ pitch sensor pixel technology is based 
on a CMOS manufacturing process with a
matrix organized in 576 rows and 1152 columns.
Based on the pitch of the pixels, the single-pixel intrinsic resolution of the Mimosa26 is expected to be $5.3\,\mm$.
A better resolution was achieved from the combination of pixels into clusters.
The sensor thresholds were optimized to improve the
intrinsic resolution. A pointing track precision of about $4\,\mm$ was achieved 
after a careful alignment of the Mimosa26 sensor planes.
The EUDET telescope provided a very precise pion-trajectory
reference for the sTGC detector. 

Event triggering was controlled by a custom Trigger Logic Unit (TLU). 
The TLU received signals from two 1 $\times$ 2\,cm$^2$ scintillators placed in front and behind the telescope. 
The TLU generated the trigger signal that was distributed to the telescope and the sTGC readout electronics.
The telescope sensors were read out in a column-parallel mode with an offset-compensated
discriminator to perform the analog-to-digital conversion which allowed a 
$115.2\,\ms$ digital binary readout. The sTGC detector was read out using the VMM1 ASIC and a custom data acquisition card.
An Arduino microcontroller board \cite{arduino} controlled the TLU system and cleared the latched trigger/signal busy provided by a custom I/O board necessary to achieve synchronization between the telescope and sTGC data.  The data of each event were sent via Ethernet using the UDP protocol and stored on a local disk array.

\begin{figure}[t!]
\centering
\includegraphics[width=\linewidth]{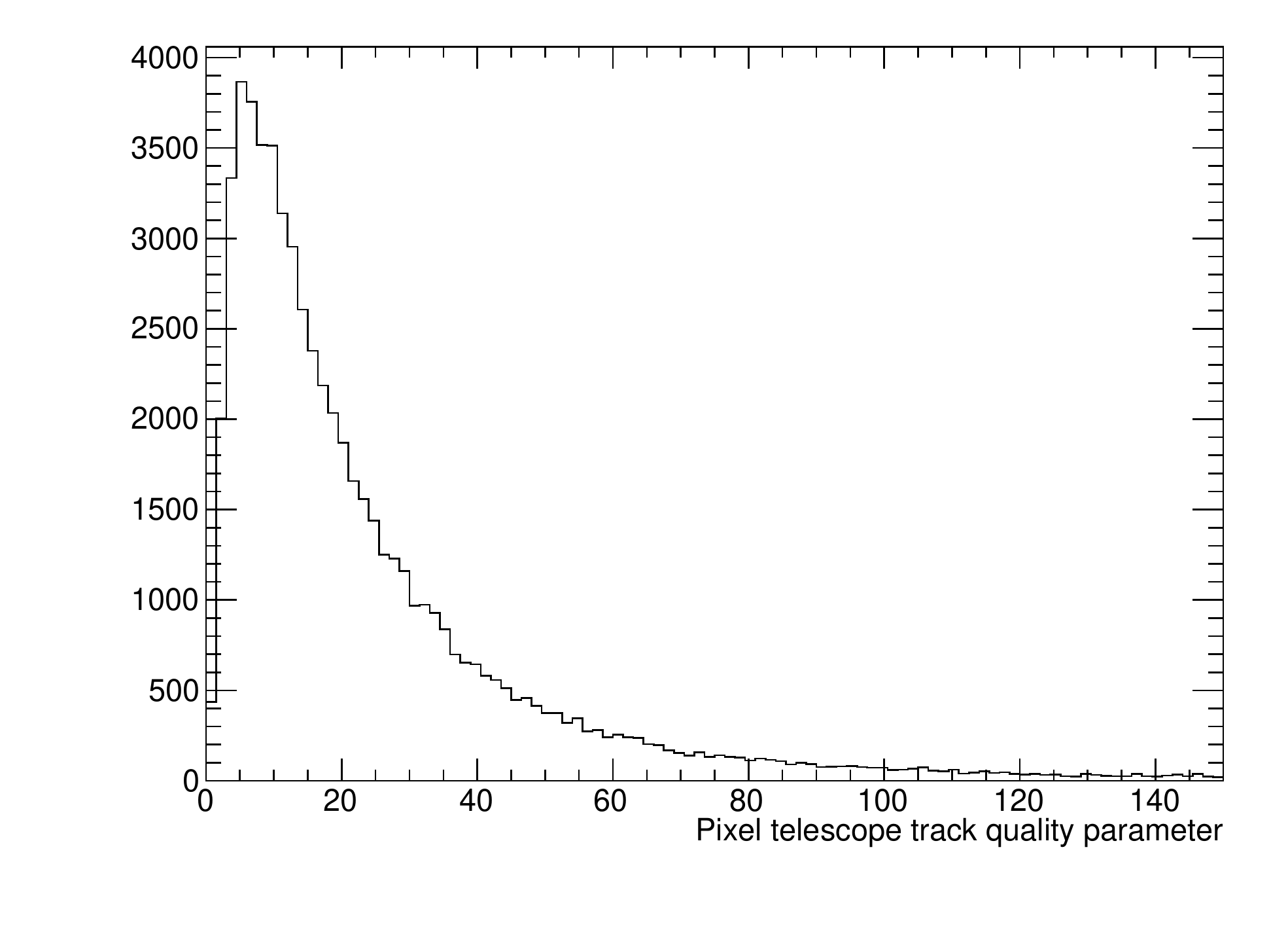}
\caption{Distribution of the $\chi^2$ track quality parameter for reconstructed tracks from the pixel telescope.}
\label{fig:track_quality}
\end{figure}

The bandwidth of the EUDET readout system and of the Arduino synchronization system
allowed to read out all six Mimosa26 sensors and the sTGC detector at a maximum possible rate of about 2\,kHz.
The silicon pixel hit positions were then used for reconstructing straight three dimensional charged-particle tracks.
A track quality parameter was obtained for each fitted pion
track based on the $\chi^2$ of the track-fit.
The distribution of track quality parameter obtained for each fitted pion track 
is depicted in Fig. \ref{fig:track_quality}. It is not perfectly distributed as a $\chi^2$ function because the errors on the hit position were only approximately Gaussian due to small random noise in the silicon pixel sensors.  
A small value of the track quality parameter 
corresponds to a straight track and a cut on this parameter 
can therefore be used to mitigate multiple scattering which are not considered in this analysis.

\subsection{Analysis Model}
\label{sec:ana}
\def\tgc{\ensuremath{\textrm{\scalefont{0.6}sTGC}}}
\def\y{\ensuremath{y_\tgc}}
\def\yo{\ensuremath{y_\mathrm{\tgc,\, 0}}}
\def\yrel{\ensuremath{\yo^\mathrm{rel}}}

\def\ypix{\ensuremath{y_\mathrm{pix}}}
\def\xpix{\ensuremath{x_\mathrm{pix}}}
\def\phixy{\ensuremath{\phi_\mathrm{xy}}}

Two analyses are performed to determine the intrinsic sTGC position resolution of a single plane. In the first analysis, the intrinsic detector resolution is estimated 
by comparing the extrapolated beam particle trajectory reconstructed by the external silicon pixel telescope, with measurements in each of the four sTGC quadruplet planes. This analysis is referred to as the 'pixel telescope analysis' hereafter. In the second analysis, the spatial resolution is estimated based on the difference between two independent measurements of the beam particle position determined in two adjacent sTGC layers
and is referred to as the 'sTGC standalone analysis'. For both analyses, the $x$-$y$ plane of the coordinate system is defined as 
the surface of the sTGC strip layer under study. The $y$-axis is defined perpendicular to the strips as shown in Fig.~\ref{fig:schematic_setup}. The sTGC strip-clusters therefore provide measurements of the particle 
position in the $y$ direction ($\y$) of this plane. 
The position resolution is directly related to the profile of induced charge on the strips. The particle position is estimated from a Gaussian fit to the induced charge 
distribution on the strips. Strip-clusters with induced charge in either 3, 4 or 5 adjacent strips are selected. 

For the pixel telescope analysis, the data from each of the four sTGC strip layers are analyzed separately.
To reduce the effect of multiple scattering on the resolution measurement, only pixel telescope tracks with track quality parameter $<$ 10 are considered.
The pixel telescope tracks provide both coordinates, $\xpix$ and $\ypix$ at the position of the sTGC layer studied.
The spatial resolution measurement is obtained by fitting the residual distribution $\y-\ypix$ with a Gaussian model.

\begin{figure}[t!]
\centering
\includegraphics[width=.85\linewidth]{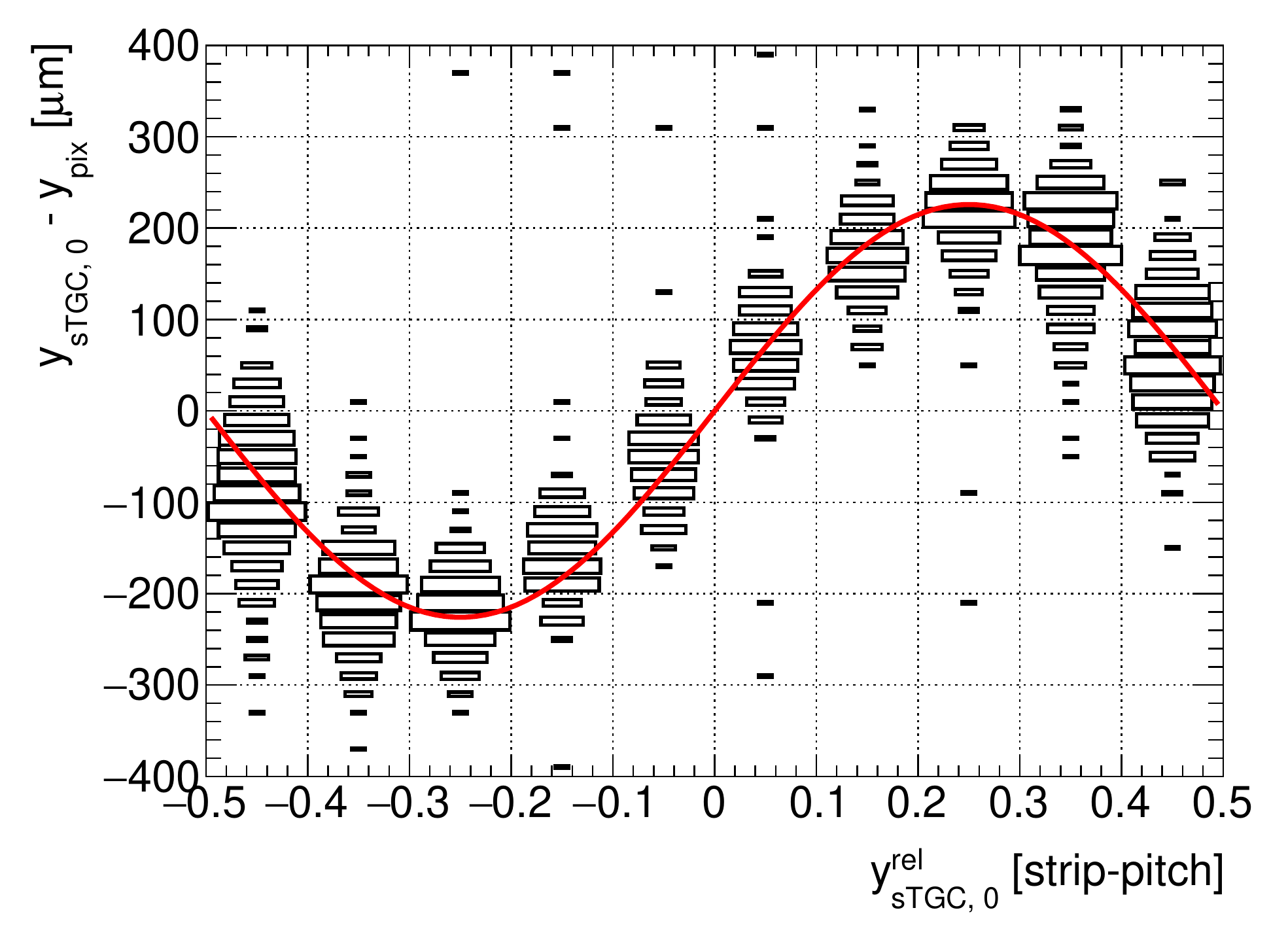}
\includegraphics[width=.85\linewidth]{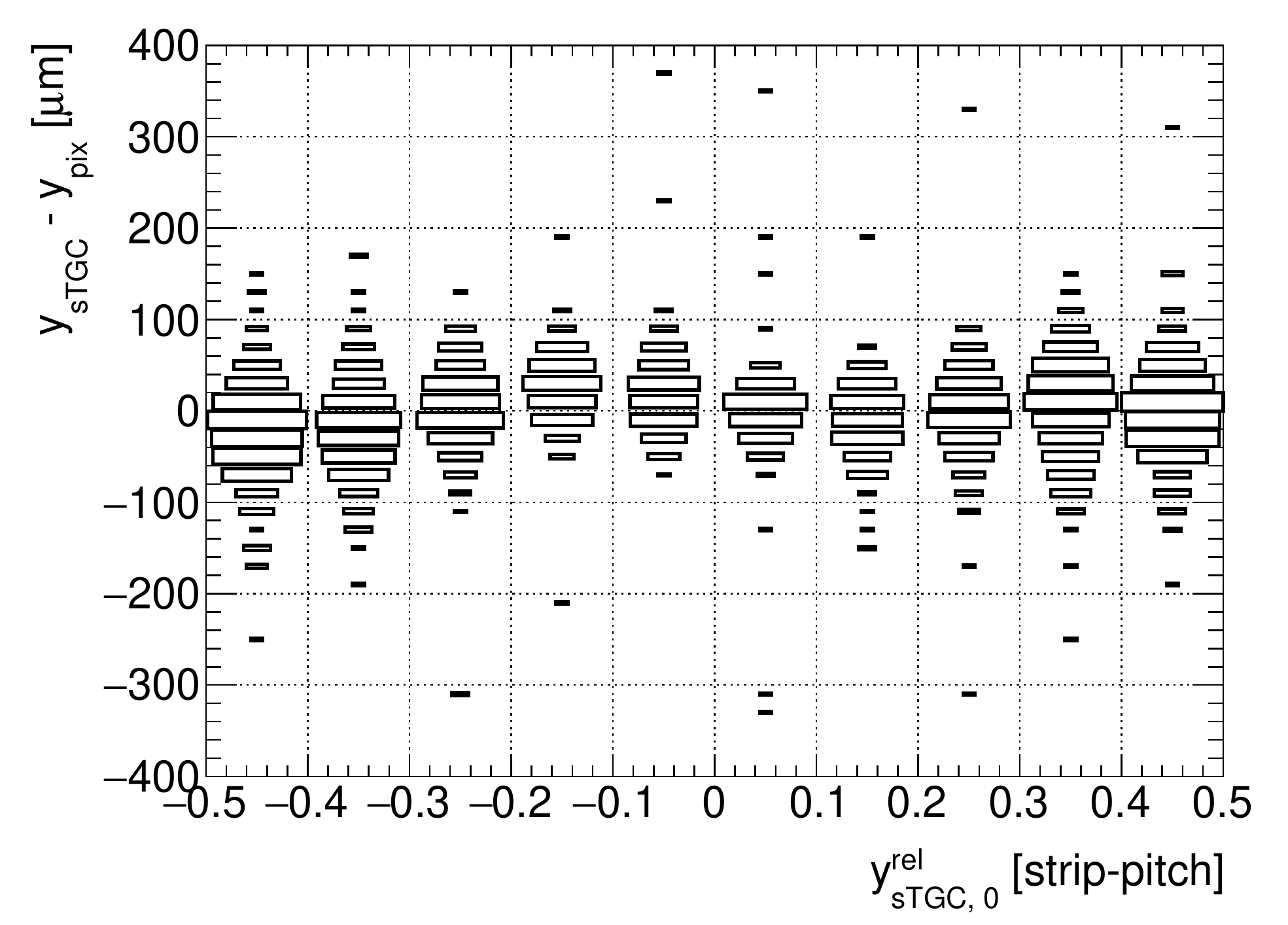}
\caption{The differential non-linearity for sTGC strip-clusters is shown before (top) and after the sinusoidal correction is applied (bottom).}
\label{fig:s-shape}
\end{figure}

The charge measured on the strips of the sTGC detector results from a spatial sampling and discretization of the induced charge.
The process of reconstructing the sTGC strip-cluster position from this sampling introduces a differential non-linearity effect on the reconstructed strip-cluster position.
The deviation of the measured strip-cluster position from the expected position (estimated by the pixel telescope track) depends on the strip-cluster position relative to the strips.
This dependence is clearly seen in the two-dimensional distributions in Fig.~\ref{fig:s-shape} (top).
It shows the $y$-residual versus strip-cluster position relative to the closest inter-strip gap center $\yrel$.
This effect is corrected using a sinusoidal function according to:
\begin{equation}
\label{eq:sshape}
\y = \yo - a_i \sin \left(\, 2\pi\,\, \yrel \, \right)
\end{equation}
where $\yo$ is the strip-cluster mean resulting from the Gaussian fit and $\y$ is the corrected particle position estimator.
The amplitude parameters are denoted $a_i$ for the 3, 4 and 5 strip-multiplicity categories; the index $i$ denotes the corresponding category.
These amplitude parameters are free parameters in the fit. 
The values of the amplitude parameters obtained from the fit to data are compatible with being equal for the three strip-cluster multiplicities as shown in Table \ref{tab:amplitudes}.
\begin{table}[t!]
\caption{Amplitude parameter $a_i$ for the differential non-linearity correction for three sTGC strip-cluster multiplicities.}
\label{tab:amplitudes}
\begin{center}
\footnotesize
\begin{tabular}{c| c}
\hline
\hline
Strip-cluster multiplicity {\it i}& Amplitude parameter $a_i$ [$\mu m$]  \\
\hline
3 & 205 $\pm$ 9  \\    	
4 & 206 $\pm$ 4  \\
5 & 211 $\pm$ 5  \\
\hline
\hline
\end{tabular}
\end{center}
\end{table}
The correction function is therefore universal and is shown in Fig.~\ref{fig:s-shape} (top).
The two-dimensional distribution after the correction is applied was found to be reasonably flat as shown in Fig.~\ref{fig:s-shape} (bottom).

\begin{figure}[b!]
\centering
\includegraphics[width=.85\linewidth]{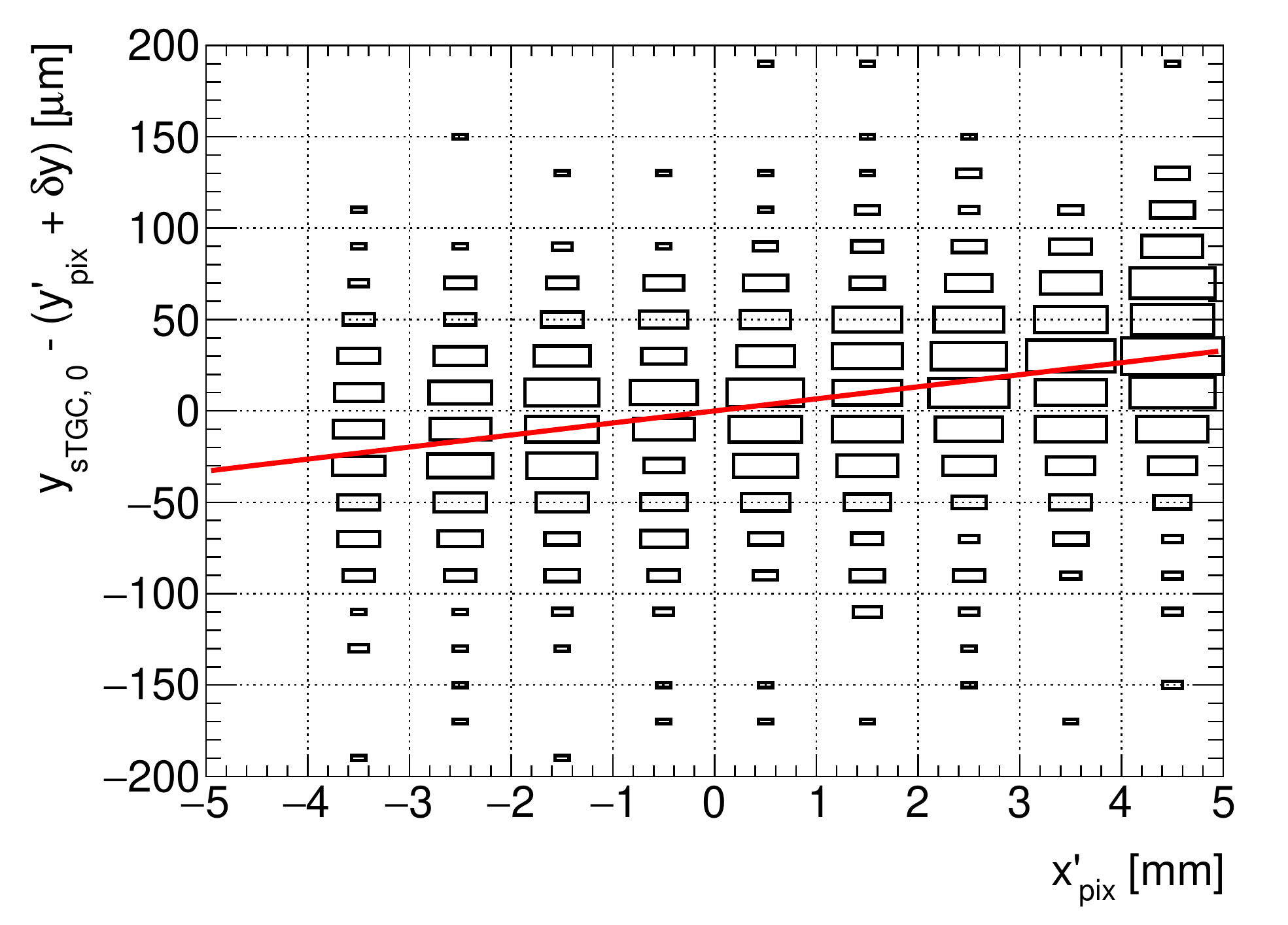}
\includegraphics[width=.85\linewidth]{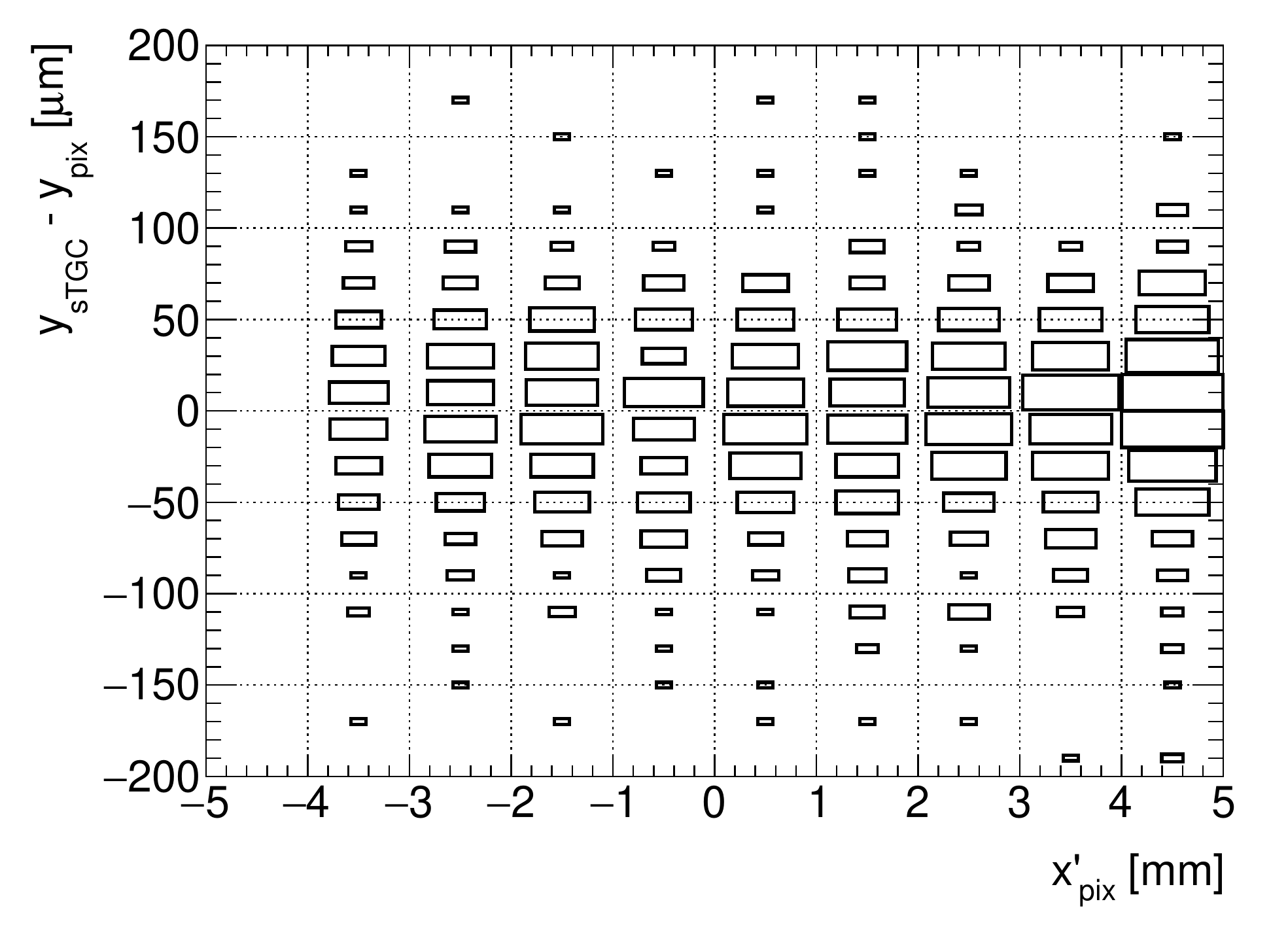}
\caption{Two-dimensional distribution of $y$-residual versus $\xpix'$ for a representative data taking run and sTGC strip-layer. For the top figure the rotation alignment correction (red line) has been omitted when computing $y$-residuals, whereas the bottom figure shows the corrected $y$-residual distribution.}
\label{fig:rel_align}
\end{figure}

The alignment of the coordinate system of the pixel telescope with respect to the  above-defined coordinate system of the sTGC layer also affects the measured residual distribution. A simple two-parameter model is used to account for translations and rotations of the two coordinate systems with respect to each other.
Both the alignment correction and the differential non-linearity correction are included {\it in-situ} in the analysis.
The alignment correction is introduced in the model by expressing the pixel track position in the sTGC-layer coordinate system $\ypix$, as a function of the track position in the pixel telescope coordinate system $\xpix'$ and $\ypix'$, and two misalignment parameters $\delta y$ and $\phixy$, as follows:
\begin{equation}
\label{eq:align}
\ypix = - \xpix' \sin{\phixy} + \ypix' \cos{\phixy}\,\, +\,\, \delta y.
\end{equation}
The variable $\delta y$ corresponds to a misalignment along the $y$-axis of the sTGC coordinate system, and $\phixy$ corresponds to a rotation of the telescope coordinate system in the $x$-$y$ plane around the $z$-axis of the sTGC coordinate system.
Translation and rotation misalignments along and around the other axes are not taken into account in this model, since they are expected to have a small impact on the determination of the intrinsic position resolution.
Fig.~\ref{fig:rel_align} shows the two-dimensional distribution of $y$-residual versus $\xpix'$ for a representative data-taking period (run) and sTGC strip-layer. In the top figure the rotation alignment correction has been omitted when computing $y$-residuals. The mean of the residual distribution increases linearly as a function of $\xpix'$, which is evidence for a small rotation between the two coordinate systems.
The red line represents the correction applied to this dataset.
Accounting for this correction results in a distribution that is independent of $\xpix'$ as shown in Fig.~\ref{fig:rel_align} (bottom).

\begin{figure}[t!]
\centering
\includegraphics[width=.9\linewidth]{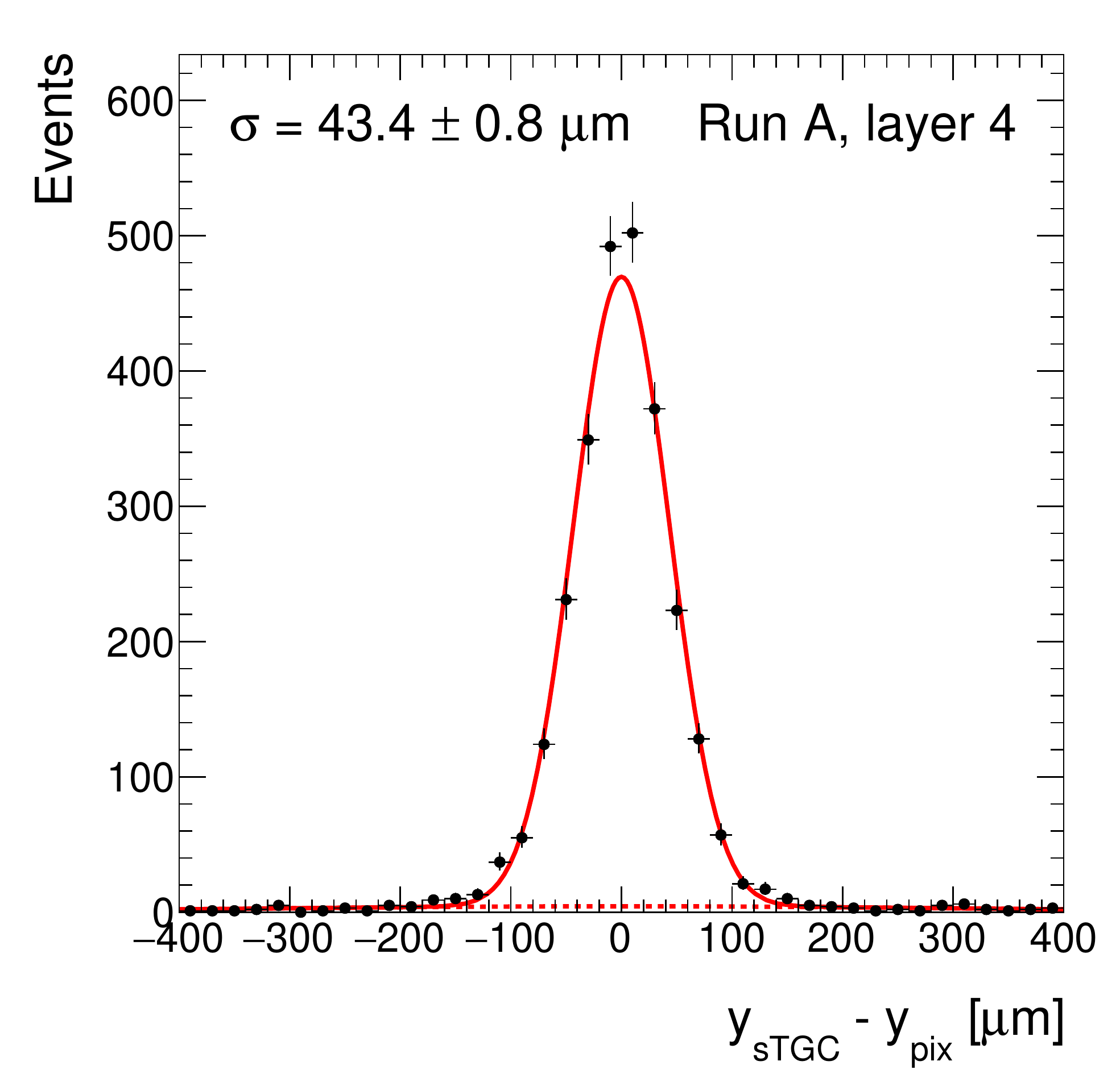}
\caption{The residual distribution after all corrections are applied together with the result for the intrinsic resolution parameter $\sigma$ for Run A, strip-layer 4.}
\label{fig:res}
\end{figure}

The global model fitted to the data is the following double Gaussian function:
\begin{eqnarray}
\label{eq:model}
\hspace{-6mm}F_{i} \hspace{-3mm} & = & \hspace{-3mm} F_{i}(\yo,\, \yrel,\, \xpix',\, \ypix' ;\, \delta y,\, \phixy,\, a_i,\, \sigma,\, f,\, \sigma_\mathrm{w}) \nonumber \\
      \hspace{-9mm} & = & \hspace{-3mm} f \, G( \y - \ypix;\, 0,\, \sigma) \,\, + \\
     \hspace{-9mm} &  & \hspace{-3mm} (1-f) \, G( \y - \ypix;\, 0,\, \sigma_\mathrm{w}); \nonumber
\end{eqnarray}
where $G$ denotes a Gaussian function.
The parameter $f$ determines the relative normalization of these two Gaussian functions.
The value of $f$ represents the fraction of the data parameterized by the narrow Gaussian and it is typically around 95\%.
The first Gaussian represents the core of the residual distribution.
The width parameter $\sigma$ is the parameter of interest for the determination of the intrinsic position resolution.
A wider Gaussian of width parameter $\sigma_\mathrm{w}$ covers reconstructed strip-clusters from background sources.
The residual distribution $\y - \ypix$ is shown in Fig.~\ref{fig:res} together with the result for the intrinsic resolution parameter $\sigma$ for a representative data taking run and sTGC strip-layer.

For the sTGC standalone analysis, the sTGC strip-cluster positions are corrected for the differential non-linearity obtained from the pixel telescope analysis.
Assuming the two sTGC strip-layers to be identical, the position resolution of a single layer $\sigma$ can be estimated from the width of the distribution of pairwise sTGC strip-layer position residuals $\sigma=\sigma_{residual}/\sqrt{2}$.

\subsection{Intrinsic Position Resolution}
\label{sec:resolution}
\begin{figure}[t!]
\centering
\includegraphics[width=0.8\linewidth]{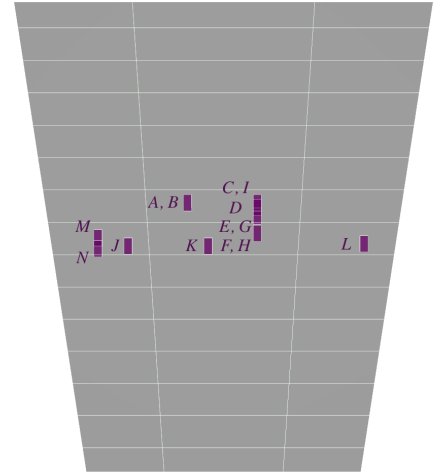}
\caption{Summary of the beam positions with respect to the sTGC detector for various data taking runs during the beam tests at Fermilab.}
\label{fig:eventdisplay}
\end{figure}

The determination of the intrinsic sTGC position resolution described in the previous section is applied to data accumulated during several data taking runs. Using the motion table on which the sTGC detector was mounted, the position of the sTGC detector was varied in the two dimensions perpendicular to the pion beam. The  beam position on the sTGC detector is summarized for various runs in Table \ref{tab:beam_positions}. Fig.~\ref{fig:eventdisplay} shows the beam positions for each run configuration. 
A large area of about 65 $\times$ 12\,cm$^2$ of the sTGC detector was exposed to the pion beam, enabling a test of the homogeneity of the detector.

\begin{figure}[b!]
\centering
\includegraphics[width=\linewidth]{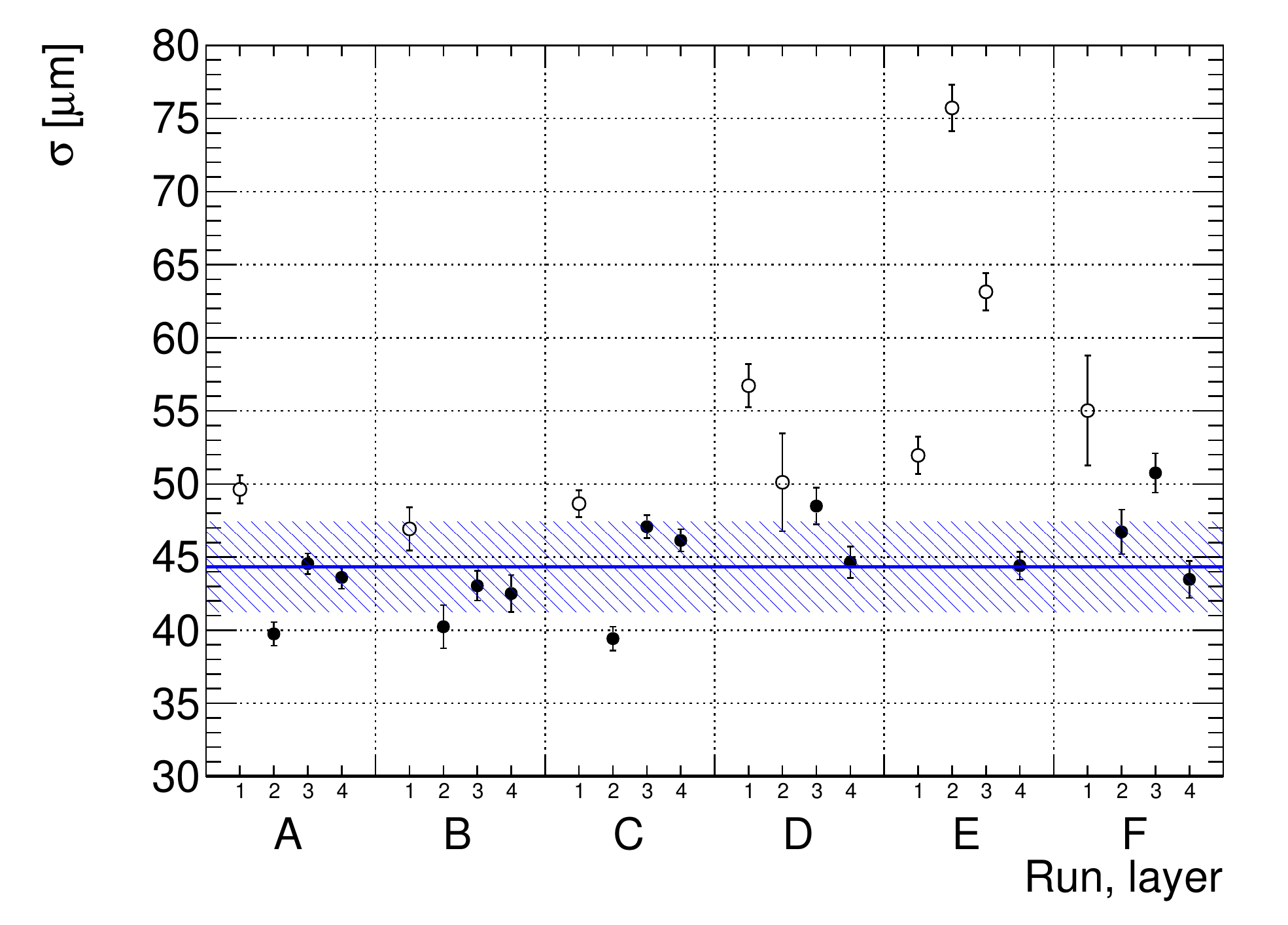}
\caption{Summary of the measured intrinsic sTGC resolution using the pixel telescope analysis for different data taking runs. The beam position on the sTGC detector for each run is summarized in Table \ref{tab:beam_positions}.  Results for runs with no expected degradation due to sTGC detector support structure or calibration are shown as black filled circles. The horizontal line represents the average resolution for these runs whereas the hashed band represents the RMS spread.
Results for the remaining runs are shown as open circles. }
\label{fig:resolution_combined}
\end{figure}

The intrinsic position resolution of each sTGC strip-layer in the sTGC quadruplet was measured using the pixel telescope analysis. For each data taking run, the same pixel telescope tracks are used to determine the intrinsic resolution in each of the four sTGC strip-layers. The results for Runs A -- F are shown in Fig.~\ref{fig:resolution_combined}. 
\begin{table}[t!]
\caption{Beam position in the coordinate system of the sTGC detector and corresponding data taking run identifier during the beam tests at Fermilab.}
\label{tab:beam_positions}
\begin{center}
\footnotesize
\begin{tabular}{c| rr}
\hline
\hline
 Run Identifier & x [mm] & y [mm]  \\
\hline
A & 2354 & 328 \\    	
B & 2354 & 328 \\	
C & 2183 	& 328 \\	
D & 2183 	& 340 \\	
E & 2183 	& 361 \\	
F & 2183 	& 404 \\	
G & 2182 	& 361 \\	
H & 2182 	& 405 \\	
I  & 2180 	& 328 \\	
J  & 2500 	& 435 \\	
K  & 2303 	& 435 \\	
L  & 1920 	& 430 \\	
M  & 2574 & 415 \\	
N  & 2574 & 442 \\	
\hline
\hline
\end{tabular}
\end{center}
\end{table}

The resolution measurements in the first sTGC strip-layer is larger for all runs because the individual readout channel baseline correction was not available. The analysis of Run E revealed that the beam crossed an internal mechanical support structure in strip-layers two (wire support) and layer three (spacer button) of the sTGC detector. The largest degradation in resolution, to about $80\,\mm$, is observed in layer-2 for Run E. The resolution measurements for configurations with expected degradation due to sTGC support structure or missing readout channel baseline correction are shown as open circles in Fig.~\ref{fig:resolution_combined}. 
All measurements with good configurations are shown as black filled circles. The average of these measurements is about $45\,\mm$ with an RMS spread of $8\,\mm$. The largest source of systematic uncertainty in this analysis is assumed to originate from multiple scattering. The magnitude is about $6\,\mm$, estimated from the dependence of the measured sTGC position resolution on the pixel telescope track quality parameter cut.
The results are well below $100\,\mm$ and therefore meet the ATLAS requirement for the NSW upgrade project. 
\begin{figure}[b!]
\centering
\includegraphics[width=.9\linewidth]{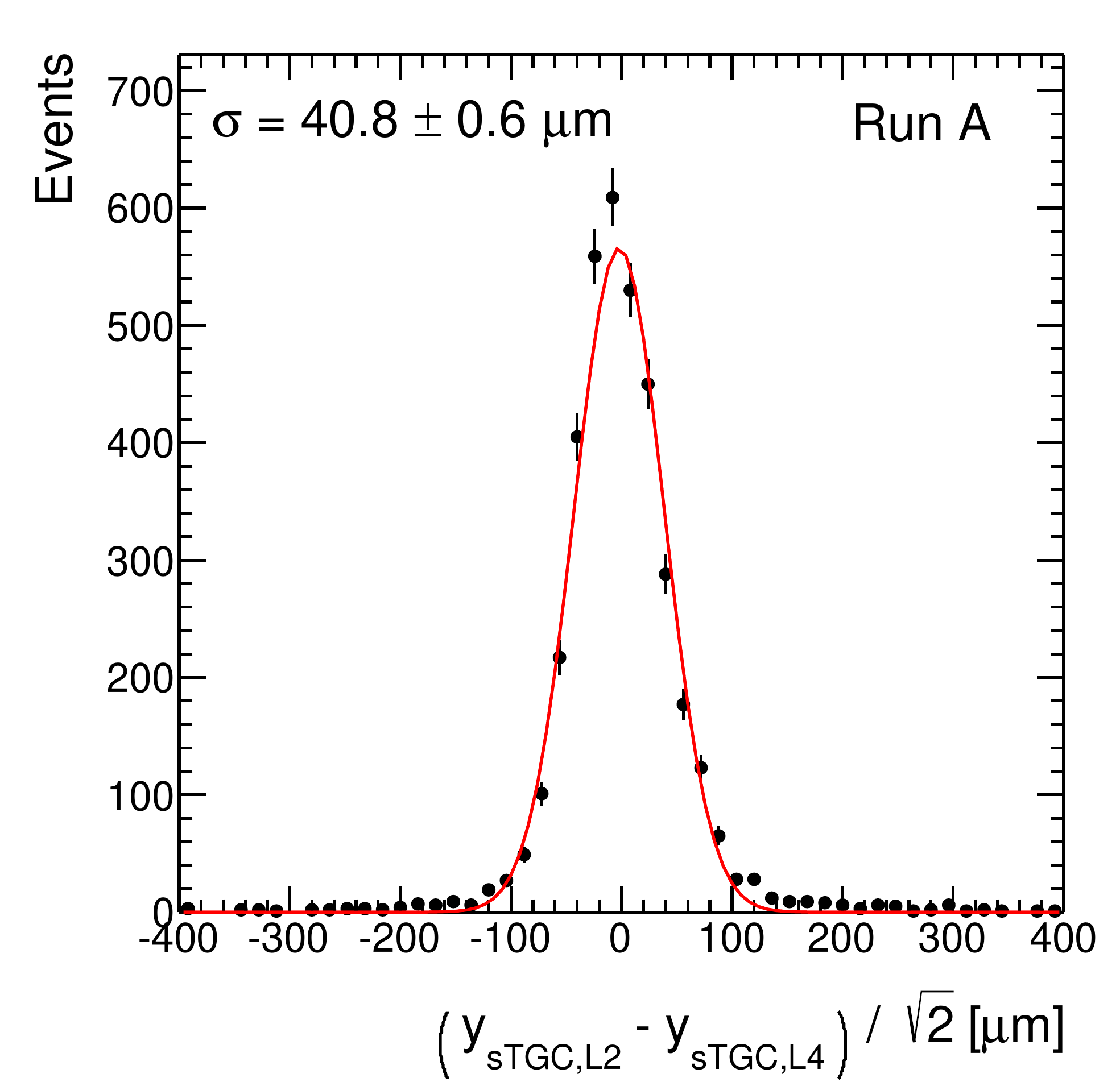}
\caption{Resolution estimate based on adjacent sTGC strip-layer position residual distributions for a representative sTGC standalone data taking run.}
\label{fig:res_pair_resolution}
\end{figure}

For Runs G -- I, no synchronized pixel telescope and sTGC data were available. 
The analysis of these runs revealed that the pion beam crossed the wire support structure in at least one of the sTGC layers of the quadruplet. Here the resolution is estimated using the sTGC standalone analysis.
The resolution is shown in Fig.~\ref{fig:res_pair_resolution} for a representative sTGC standalone run. 
\begin{figure}[t!]
\centering
\includegraphics[width=\linewidth]{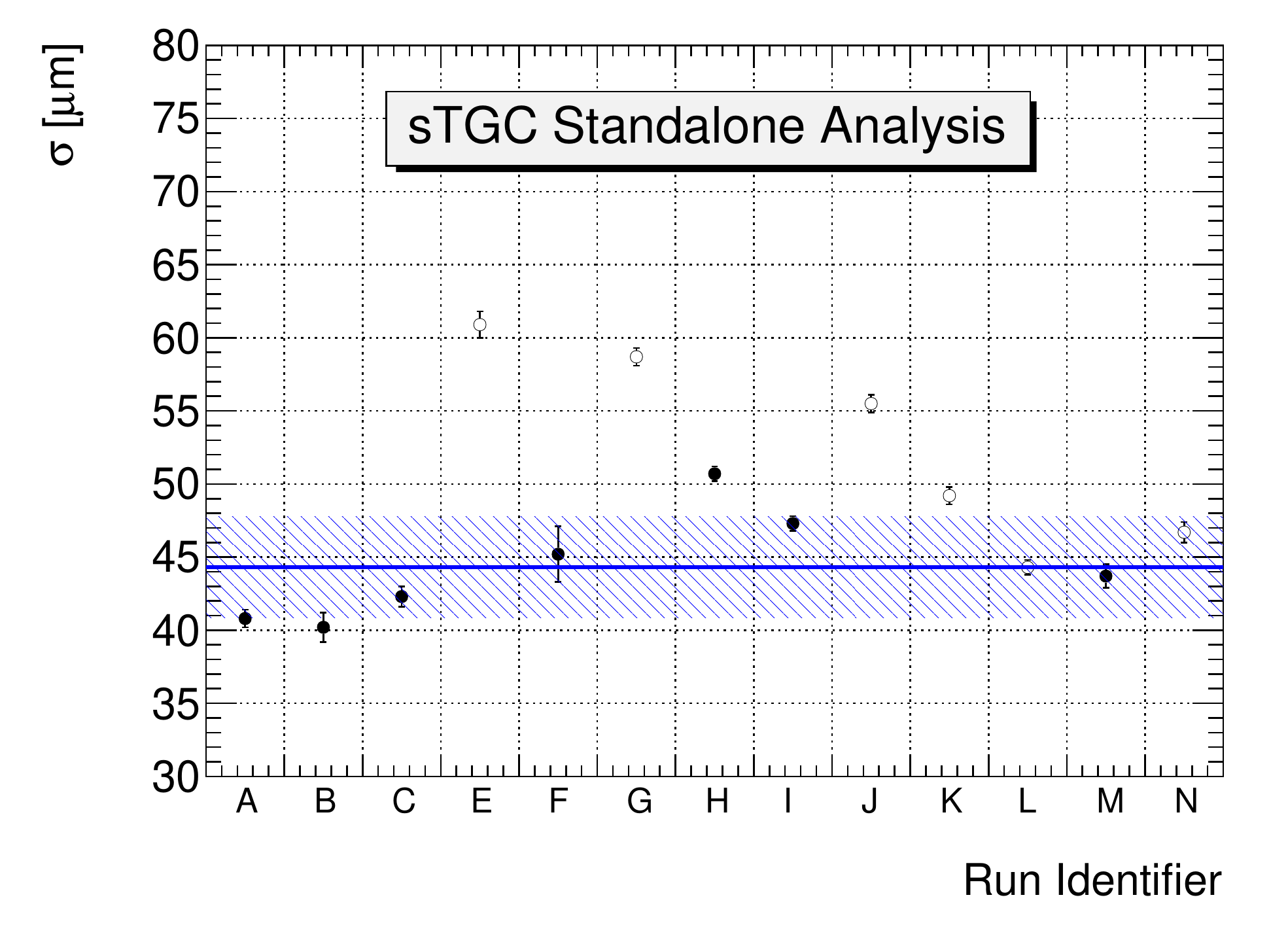}
\caption{Summary of the measured intrinsic sTGC resolution using the sTGC standalone analysis for different data taking runs. Results for runs with no expected degradation due to sTGC detector support structure or calibration are shown as black filled circles. The horizontal line represents the average resolution for these runs whereas the hashed band represents the RMS spread. Results for the remaining runs are shown as open circles.}
\label{fig:resolution_summary}
\end{figure}
A summary of the measurements for all runs is shown in Fig.~\ref{fig:resolution_summary}. Measurements for configurations with expected degradation due to sTGC detector support structure or calibration are shown as open circles. The measurement with no expected degradation are shown as black filled circles. The weighted average of these measurements is also about $45\,\mm$,  consistent with the pixel telescope analysis, with a spread of $8\,\mm$.

\section{Pad Readout Measurements at CERN}
\label{sec:cern}

\begin{figure}[t!]
\centering
\includegraphics[width=0.7\linewidth]{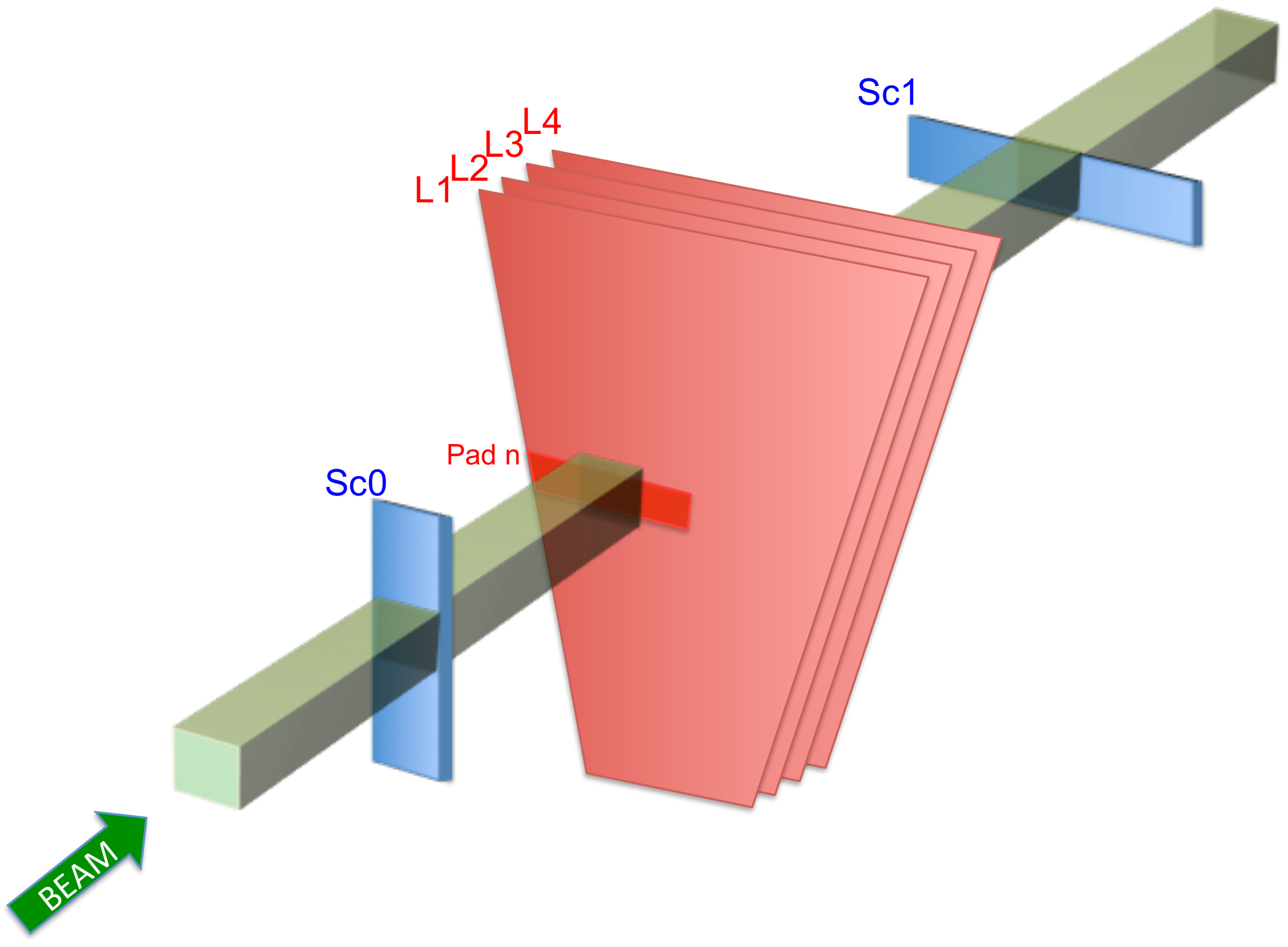}
\caption{Schematics of the experimental setup of the CERN beam test experiment. Two scintillators are positioned perpendicular to each other, one before and one after the chamber, which define a coincidence area of 12 $\times$ 12~cm$^2$, centered on pad $n$.}
\label{fig:CERN_exp_setup}
\end{figure}

A   test beam experiment was conducted at the CERN H6 facilities, using a $130\, \mathrm{GeV}$ muon beam of about 4\,cm radius, to test the characteristics of the pads.
The setup is shown in Fig.~\ref{fig:CERN_exp_setup}. The system was triggered by a set of scintillators with a 12 $\times$ 12\,cm$^2$ coincidence area.

\subsection{Deadtime and efficiency measurements}

The preliminary front-end electronics based on the VMM1 ASIC is not adapted to the long time drifting of the late clusters in the sTGC detector (change of baseline).
This leads to a large dead time in its response, which in turn leads to an inefficiency of the system when running at high rate (typically 80 -- 90\% efficiency at 100\,Hz/cm$^2$).
%
To ensure that no inefficiency was due to the detector itself, the large cathode pads were used to estimate the detector efficiency,
which was measured by looking at the analog output of the front-end amplifier.
The efficiency of pad $n$ in the first layer was defined with respect to the coincidence of the trigger with a signal in the fully overlapping pad of the second layer.

Figure~\ref{fig:CERN_scope} shows two typical analog pulses:
\begin{figure}[b!]
\centering
\includegraphics[width=0.8\linewidth]{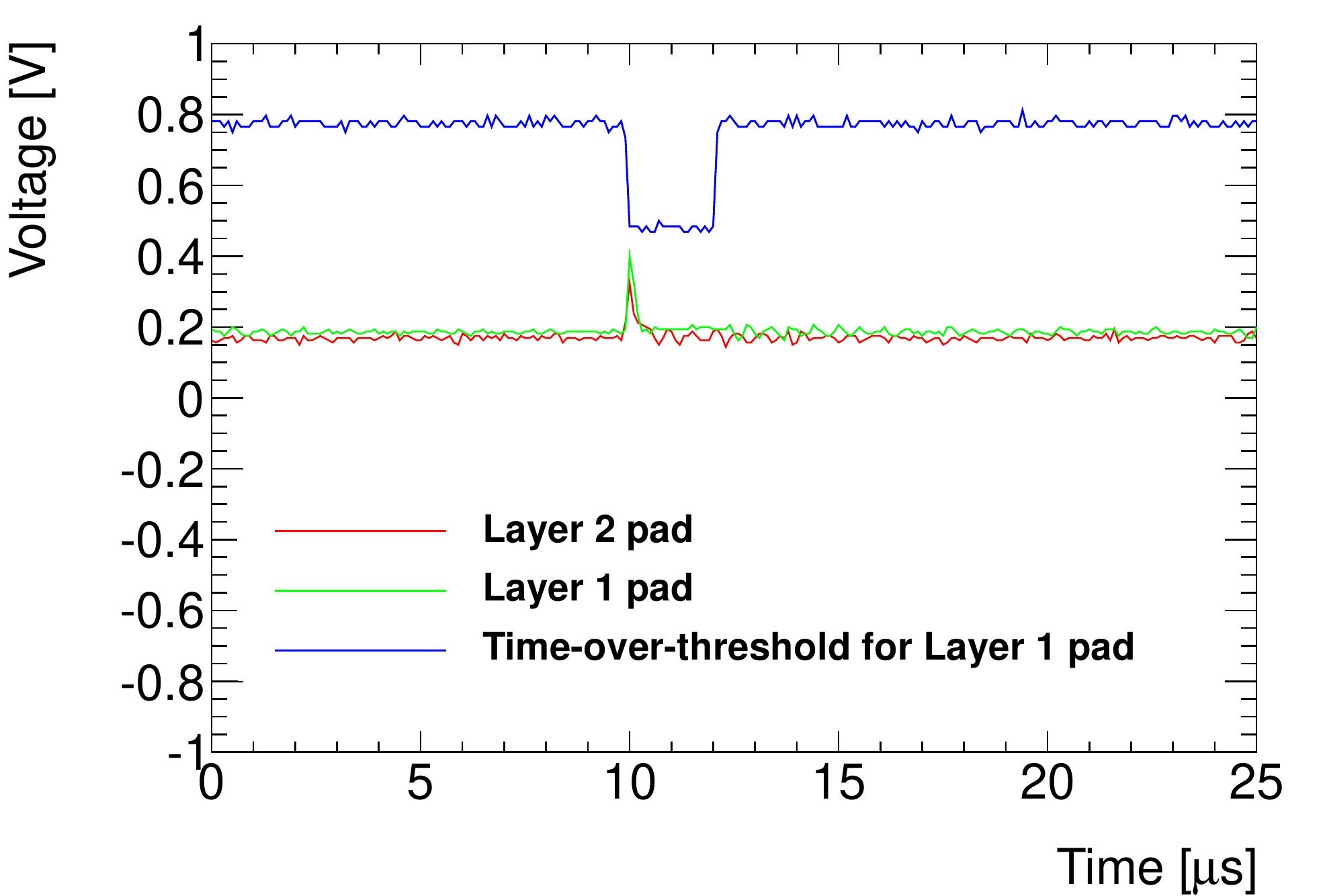}
\includegraphics[width=0.8\linewidth]{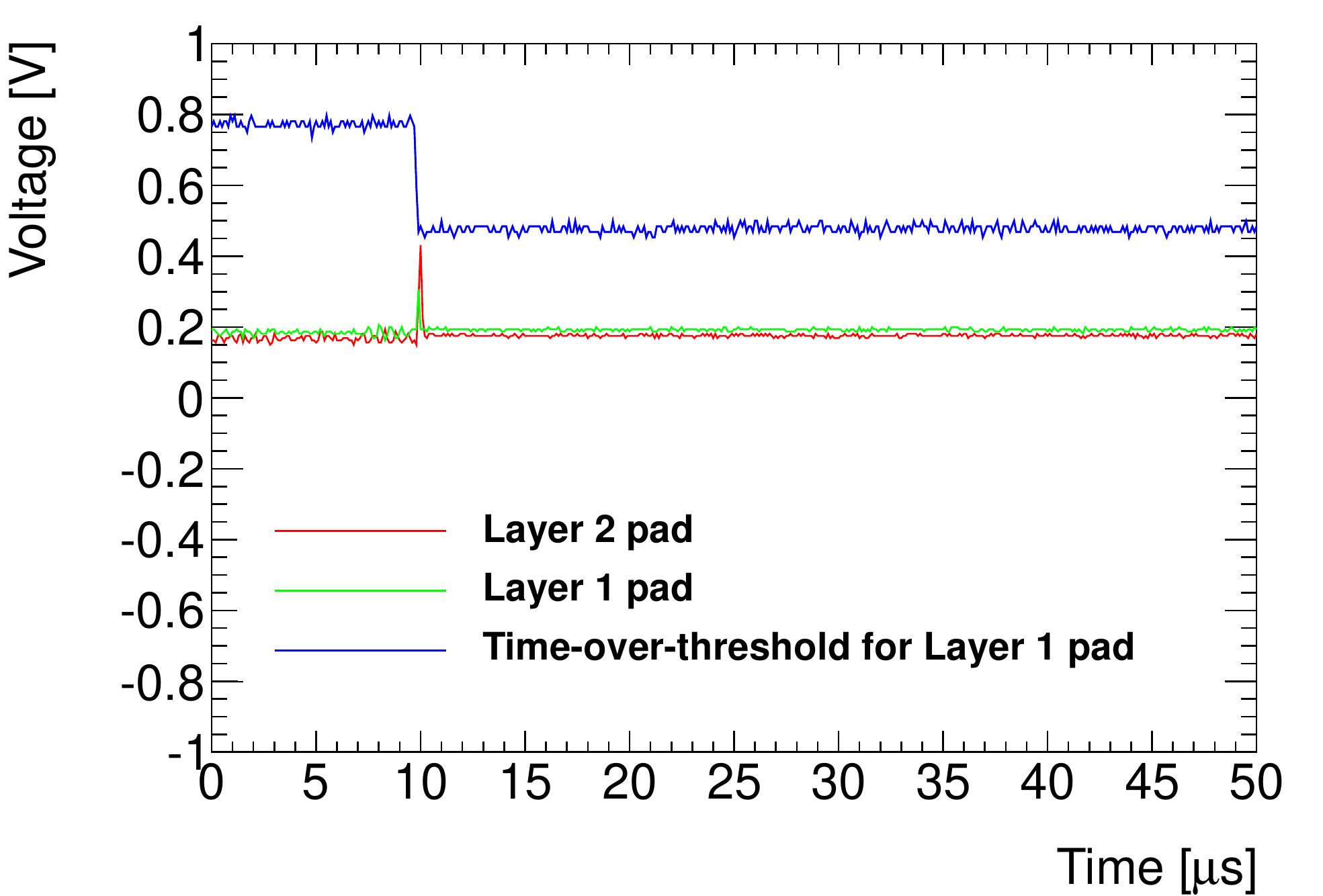}
\caption{ Oscilloscope event examples, where the amplifier shows a short dead time (top) and where the dead time lasted for tens of microseconds (bottom). }
\label{fig:CERN_scope}
\end{figure}
one where the amplifier shows a short dead time following the detector signal (top),
and one where the dead time lasted for tens of micro-seconds (bottom).
By taking hundreds of triggers and checking if the detector signal was within the live-part of the front end, 
it was determined that the detector was 100\% efficient. 

Furthermore, by placing the beam and a scintillator coincidence triggering area of 12 $\times$ 1\,cm$^2$ centered on a single 8 $\times$ 60\,cm$^2$ pad and looking at the signal in neighbouring pads, 
one could determine that any cross-talk to neighbouring pads does not exceed 5\% of the cases.
This should be considered as an upper limit since the muon beam did contain also two muons per event, where the second muon could be outside the region that was instrumented with strip-readout electronics.

\subsection{Charge sharing between pads} 

\begin{figure}[h!]
\centering
\includegraphics[width=0.7\linewidth]{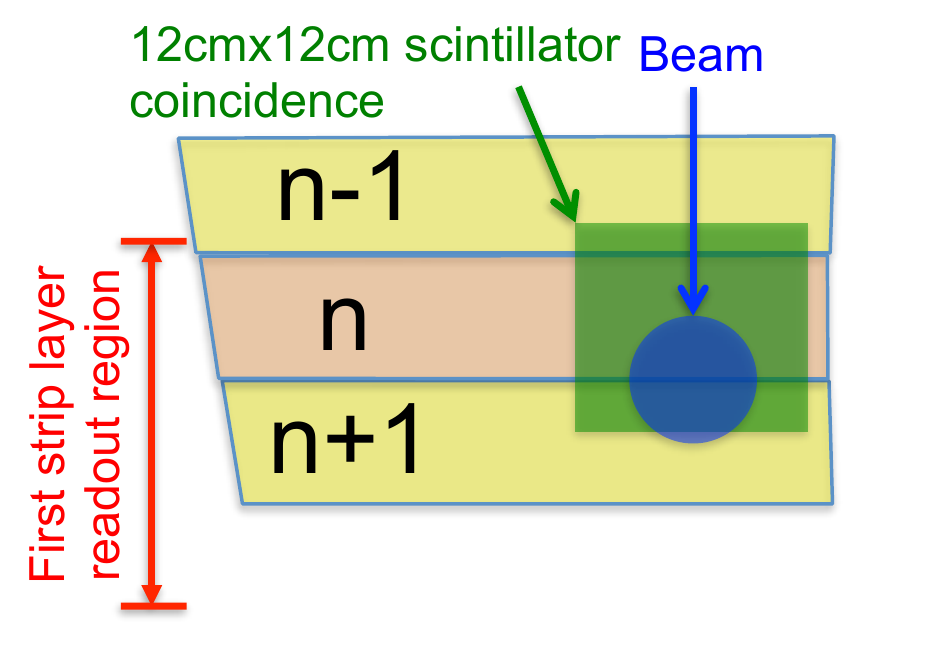}
\caption{Schematics of the experimental setup for charge sharing measurements. The beam and the scintillator coincidence area cover the transition between pad $n$ and pad $n+1$.}
\label{fig:CERN_exp_setup_2Nov}
\end{figure}

To study the transition region between pads, the scintillator coincidence triggering area and the particle beam were centered
between pad $n$ and pad $n+1$ of the first layer, as illustrated in Fig.~\ref{fig:CERN_exp_setup_2Nov}.

\begin{figure}[h!]
\centering
\includegraphics[width=0.7\linewidth]{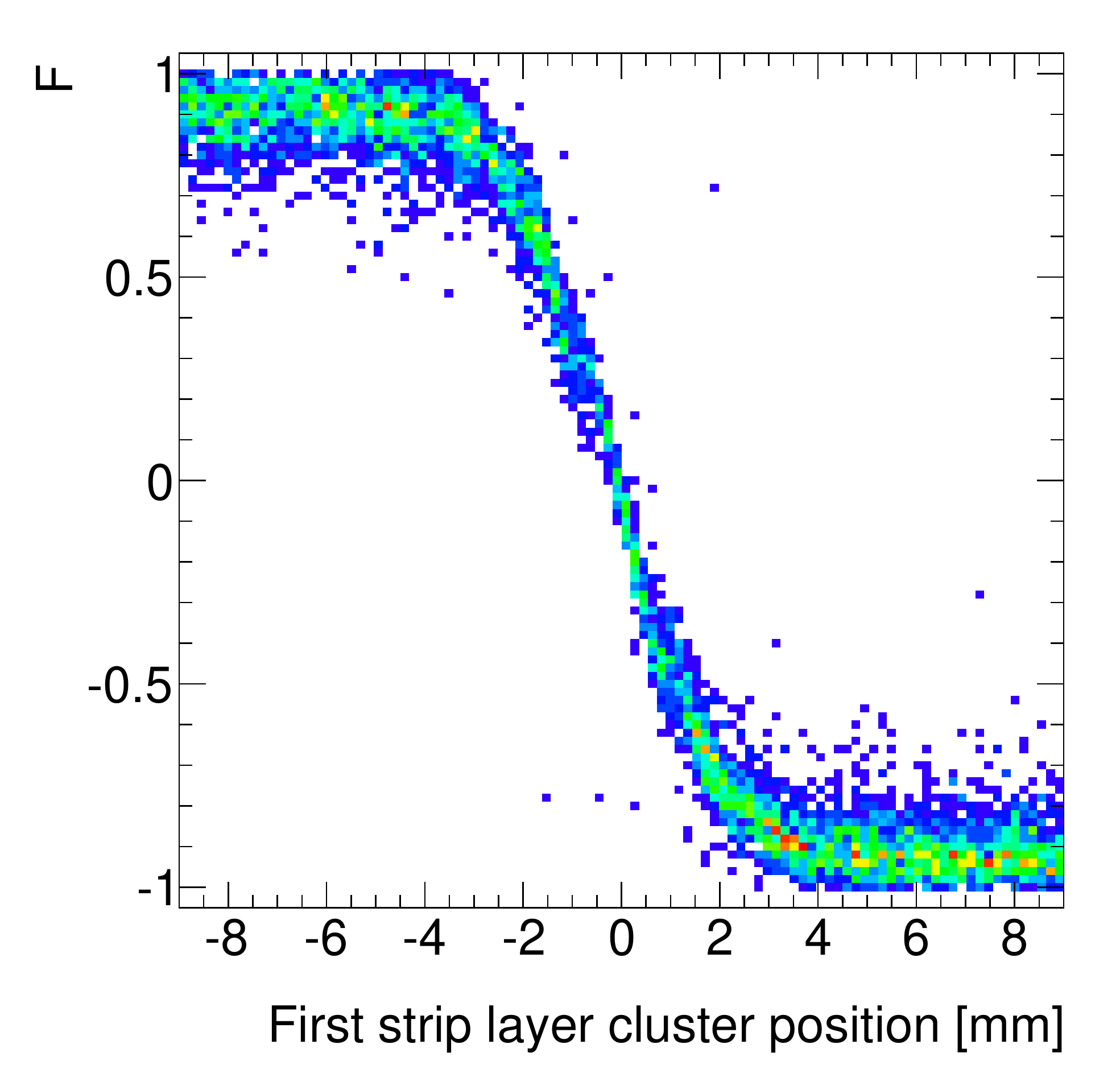}
\caption{Fraction of the charge collected by pad $n$ as a function of the position with respect to the center of the transition region. F=1 implies all the charge is deposited in pad $n$, while F=-1 means all the charge goes to pad $n+1$.}
\label{fig:CERN_pad_charge_sharing}
\end{figure}

After applying timing quality requirements on the strip and pad hits, the channel baseline values are subtracted from the analog peak values.
Strip-clusters with induced charge in either 3, 4 or 5 adjacent strips are selected and calibrated in the same way as for the Fermilab beam test. 
Events with a single strip-cluster in the first layer and the second layer are selected. 
The strip-cluster position (mean of the fitted gaussian) in the first layer is used to define the position of the particle going through the detector.
The events are further required to contain a hit above threshold on either pad $n$ or pad $n+1$. 
The charge fraction (F) is defined using the analog peak values ( P ) of the two adjacent pads:
\begin{equation}
F=\frac{P_{n} - P_{n+1}}{P_{n} + P_{n+1}}
\end{equation}
Fig.~\ref{fig:CERN_pad_charge_sharing} shows the charge fraction as a function of the position with respect to the center of the transition region between the pads.
It shows that the transition region, where the two pads share more than 70\% of the induced charge, spans about 4\,mm.

\section{Conclusions}
The spatial resolution for a full-size sTGC prototype detector for the ATLAS NSW upgrade has been measured in a $32\, \mathrm{GeV}$ pion beam test experiment at Fermilab. A six-layer silicon pixel telescope has been employed to characterize the sTGC detector and to correct for differential non-linearity of the reconstructed sTGC strip-cluster position. At perpendicular incidence angle, single strip-layer position resolutions of better than $50\,\mm$ have been obtained, uniform along the sTGC strip and perpendicular wire directions, well within design requirements. The characteristics of the sTGC pad readout have been measured in a $130\, \mathrm{GeV}$ muon beam test at CERN. The transition region between readout pads has been found to be 4\,mm, and the pads have been found to be fully efficient.

\section{Acknowledgements}
The authors would like to thank the members of the NSW collaboration for their contributions and in particular the NSW electronics group for providing the VMM1 based readout system. We would also like to thank B. Iankovski, F. Balahsan, G. Cohen, S. Sbistalnik, A. Klier from the Weizmann Institute of Science, M. B. Moshe from Tel-Aviv University, M. Batygov, M. Bowcock, Y. Baribeau, P. Gravelle from Carleton University, R. Openshaw from TRIUMF and J. Fried from BNL for supporting this work. We are also thankful to the firm MDT SRL (Milan) for their efforts in producing the first prototypes of large cathode boards, which made the construction of the tested module possible. We acknowledge the support of of the Israeli Science Foundation (ISF), CIRP-Israel-China Collaboration Grant 549/13, the MINERVA foundation - project 711143, the Benoziyo Center for Particle Physics, the Natural Sciences and the Engineering Research Council (NSERC) of Canada and CONICYT, Chile.




\bibliographystyle{elsarticle-num}
\bibliography{<your-bib-database>}



\end{document}